# Artificial Creations: Ascription, Ownership, Time-Specific Monopolies

Raj Shekhar

Institute of Public Policy, National Law School of India University, Bengaluru

## Abstract

Creativity has always been synonymous with humans. No other living species could boast of creativity as humans could. Even the smartest computers thrived only on the ingenious imaginations of its coders. However, that is steadily changing with highly advanced artificially intelligent systems that demonstrate incredible capabilities to autonomously (i.e., with minimal or no human input) produce creative products that would ordinarily deserve intellectual property status if created by a human. These systems could be called "artificial creators" and their creative products "artificial creations". The use of artificial creators is likely to become a part of mainstream production practices in the creative and innovation industries sooner than we realize. When they do, intellectual property regimes (that are inherently designed to reward human creativity) must be sufficiently prepared to aptly respond to the phenomenon of what could be called "artificial creativity". Needless to say, any such response must be guided by considerations of public welfare. This paper analyzes what that response ought to look like by revisiting the determinants of intellectual property and critiquing its nature and modes. This understanding of intellectual property is then applied to investigate the determinants of intellectual property in artificial creations so as to determine the intrinsic justifications for intellectual property rewards for artificial creativity, and accordingly, develop general modalities for granting intellectual property status to artificial creations. Finally, the treatment of "artificial works" (i.e., copyrightable artificial creations) and "artificial inventions" (i.e., patentable artificial creations) by current intellectual property regimes is critiqued, and specific modalities for granting intellectual property status to artificial works and artificial inventions are developed.

# 1. Introduction

Intellectual property is a product of the mind (World Intellectual Property Organization: 2; Christie 2011: 6). But, the mind is not an axiom. The mind is usually deemed synonymous with humans. But, non-human living entities like plants and animals have demonstrated some form of "mind-ness" that is either unique or similar to that of humans (Dawkins 2012[1]; Chauncey 2017[2]). Moreover, think about computers. Computers have been undertaking intellectual tasks that were associated exclusively with the human mind before their advent, and therefore, they could be argued to possess a form of mind-ness when they seamlessly perform complex calculations, systematically store useful information, and assist humans in composing artistic works and writing scientific papers. With our current understanding of the mind, we cannot agree (at least not entirely) on who possesses it, what constitutes it, and how does it function. Hence, as the "axiomatic mind" does not exist, we must acknowledge the mind-ness of all living entities – human or non-human – in all their forms and variety.

The absence of an "axiomatic mind" rarely obstructed the development and practice of intellectual property regimes around the world.[3] Intellectual property was safely presumed to emanate exclusively from the exercise of the highest form of intellectual faculty possessed only by humans, i.e., creativity. Creativity in intellectual property regimes was understood as the unique ability in humans to create objects of art and innovation that could enrich human culture and advance human development – "creative products". These creative products were generally deemed original and valuable by intellectual property regimes – reassuringly, not quite varying from the broad philosophical understanding of creativity (Gaut 2010: 1039).

Intellectual property regimes granted the status of intellectual property to these creative products. The intellectual property status granted to these creative products included a set of rewards, namely, ascription – acknowledging the creator, ownership – of a set of exclusive rights over the creation, and time-specific monopolies – setting the expiry date for those exclusive rights. These intellectual property rewards were intended to encourage and inspire production of more creative products for the continuous enrichment of human culture and advancement of human development. Since, human creativity could not be mimicked or surpassed by any other entity – living (plants and animals) or non-living (computers) – producing creative products was the exclusive domain of human creativity — and therefore, humans remained the undisputed recipients of intellectual property rewards.

---

[1] Exploring the possibilities of consciousness in non-human animals.

[2] Claiming that plants do have minds by virtue of being living entities.

[3] *See* Naruto v. Slater No. 16-15469, 2018 WL 1902414 (9th Cir. Apr. 23, 2018) – the decision apparently held that *Naruto* (the macaque monkey) could not hold and enforce copyright.

It is true that computers could perform intellectual tasks and demonstrated mind-ness, but their capacity and function were limited by their underlying code (or their coder's creativity). Hence, the computer's mind was completely possessed and constituted by the coder, and it functioned to ultimately serve the coder's will. More importantly, it could not demonstrate the capability to autonomously produce creative products deserving of the intellectual property status. Simply put, computers were not creative.

Artificial creators are geared towards changing just that by seeking not to merely assist but replace human creators and, in many ways, challenge human creativity. These highly advanced artificially intelligent systems (Russell and Norvig 2013; Kaplan 2016)[4] are capable of producing various forms of creative products deserving of the intellectual property status autonomously (i.e., with minimal or no human input) — "artificial creations". Such are the claims of the developers of these systems and their proponents who are convinced that creativity has ceased to be synonymous with humans (Keats 2006[5], Abbott 2016a: 19; Karpathy 2015[6]; The Next Rembrandt 2016[7]; Veale and Cardoso 2019; Imaginations Engines Inc. a[8], b[9]; AIVA[10]). This technological breakthrough culminating in the phenomenon of artificial creativity — artificial creators producing artificial creations autonomously — is destined to have an enormous bearing on intellectual property regimes that seek to reward creativity.

---

[4] No universal definition of an artificially intelligent system exists. Artificially intelligent technologies could be understood as a constellation of technologies that are capable of mimicking human brain processes and action.

[5] *The Invention Machine* (an artificially intelligent system built using genetic programming) could produce patentable creations, including antennae, circuits, and lenses.

[6] An artificially intelligent system built on artificial neural networks could compose texts (copyrightable creations) closely resembling the iconic style of William Shakespeare, the great English playwright of the 16th century.

[7] An artificially intelligent system based on facial-recognition technology could produce paintings (copyrightable creations) closely resembling to that of Rembrandt, the legendary Dutch painter.

[8] *The Creativity Machine* (an artificially intelligent system built on artificial neural networks) could produce 11,000 songs over a weekend (copyrightable creations).

[9] *DABUS* or *Device for the Autonomous Bootstrapping of Unified Sentience* (a special type of Creativity Machine) could produce patentable creations, including a specially designed container-lid for robotic gripping and an emergency flashlight.

[10] An artificially intelligent system built on artificial neural networks could produce soundtracks for films, video games, and commercials by learning from the compositions of legendary artists like Beethoven, Mozart, and Bach.

The existing intellectual property jurisprudence was developed by law-makers, courts, and jurists without anticipating the possibility of artificial creativity (Vertinsky and Rice 2002: 576[11]; Ramalho 2017: 13). Even today, the legal position on artificial creativity remains unresolved (Iglesias et al. 2019[12]). National and international intellectual property regulators have begun to take official cognizance of the jurisprudential gaps around artificial creativity as it steadily makes its way to the creative and innovation industries (World Intellectual Property Organization 2019a; United States Patent and Trademark Office 2019). Broadly, regulators have the following legal options to consider:

(1) Intellectual property regimes do not discriminate between artificial creations and human creations (Abbott 2016a; Jehan 2019)[13];

(2) Intellectual property regimes treat artificial creations differently from human creations (Schönberger 2018[14]; McLaughlin 2018[15]; Lauber-Rönsberg and Hetmank 2019[16]; United Kingdom Intellectual Property Office Formalities Manual 2019[17]; European Patent Office 2020[18]; Nurton 2020[19]; Tapscott 2020[20]).

(3) Artificial creations are deemed to exceed the purview of intellectual property regimes, implying that artificial creations ought not to be granted any form of intellectual property status (Ravid and Liu 2018[21]).

---

[11] Arguing that current patent law regimes are not equipped to accommodate artificial inventions.

[12] Explaining the need for further economic and legal research that should precede the creation of any regime for the protection of artificial creativity.

[13] Arguing that artificial creators should be deemed as authors and inventors of artificial creations.

[14] Arguing that equal treatment of artificial works and works of pure human ingenuity could possibly destroy incentives for human creators

[15] Proposing that artificial works should be unpatentable.

[16] Claiming that the difficulty in implementing mandatory/voluntary disclosure of artificial creativity in copyrightable creations would lead to the integration of artificial works in the copyright regimes, disrupting the foundational doctrines of copyright.

[17] Precluding artificial inventorship for artificial creators.

[18] Rejecting the proposition of artificial inventorship by holding that the inventor must be a natural person.

[19] Reporting European Patent Office's (EPO) and United Kingdom Intellectual Property Office's (UKIPO) rejection of patent applications that had named an artificial creator as an inventor; both EPO and UKIPO held that the inventor must be a natural person.

[20] Reporting United States Patent and Trademark Office's (USPTO) rejection of patent applications that had named an artificial creator as an inventor; USPTO held that the inventor must be a natural person.

[21] Arguing that intellectual property status is irrelevant for artificial inventions.

This paper contains (a) a roadmap to arrive at the option which could demonstrate good congruence with the intrinsic justifications for intellectual property, and (b) suggest or devise intellectual property modalities accordingly which could aptly respond to artificial creativity.

## 1.1 Policy Relevance of the Paper

Artificial creators demonstrate incredible capabilities to mimic, and even surpass human creativity – that puts them on a trajectory to redefine the mainstream production practices in the creative and innovation industries. Hence, their cultural and economic impacts on our lives must not be underestimated.

Live every new technology, they promise rich benefits, but not without significant risks. Their autonomy is fascinating, but artificial, at the end of the day. They ultimately derive their autonomy from us. That means they are not entirely beyond human control.

Hence, we should assume at least some form of responsibility for (a) how we encourage and facilitate their development, deployment, and use, (b) what they ending up creating, and (c) how their artificial creations are deployed. These questions constitute the subject of policy research and policy analyses that should not be relegated to artificially intelligent systems even if it could be – if we are to retain agency over our lives.

With the likely increase in the usage of artificial creators in the creative and innovation industries, artificial creations would proliferate. If artificial creations ought to be granted the status of intellectual property in some form, the need to strike an optimal balance between private monopoly over these creations and public access to these creations (like for creative products) cannot be overemphasized – towards securing and maximizing public welfare.

Proposals from this paper could guide policymakers in developing an intellectual property regime for artificial creations in addressing this emerging critical need.

## 1.2 Structure of the Paper

In **Chapter 2**, intellectual property is introduced as a set of rewards granted by the State on an exclusive-basis to a creator for a limited period of time. Then, the factors that determine these intellectual property rewards are analyzed.

In **Chapter 3**, intellectual property rewards are discussed as incentives for human creativity that must be used with sufficient caution.

In **Chapter 4**, modes of intellectual property rewards are analyzed in terms of how ex ante determination of certain intellectual property rewards that are determined by constant factors is superior to their ex post determination, and how ex post determination of certain intellectual property rewards that are determined by variable factors is superior to their ex ante determination.

In **Chapter 5**, factors determining intellectual property rewards (or intrinsic justifications for intellectual property) are examined and analyzed in artificial creations.

In **Chapter 6**, proposals for general modalities for artificial creations are developed.

In **Chapter 7**, based on the findings in the preceding chapters, the treatment of artificial creativity by current national intellectual property laws are critiqued.

In **Chapter 8**, a blueprint (comprising specific modalities) for granting intellectual property status to artificial works and artificial inventions is proposed.

In **Chapter 9**, key findings and recommendations from this paper are summarized.

## 2. Determinants of Intellectual Property

Intellectual property is a set of rewards granted by the State over creative products on an exclusive-basis to the creator for a limited period of time to secure and maximize public welfare. These rewards include: ascription – acknowledging the creator, ownership – of a set of exclusive rights over the creation, and time-specific monopolies – setting the expiry date for those exclusive rights. These rewards are determined by the following factors:

### 2.1 Human Ingenuity in Creative Products

Intellectual property regimes treat creativity as an intellectual function, i.e., an act whose origins lie in the human mind (Christie 2011: 6). However, not all products emanating from the intellectual function are treated as creative by intellectual property regimes.

It is only when the intellectual function ends up creating a product that demonstrates a modicum of human ingenuity, do intellectual property regimes start recognizing that product as creative – with the distinct purpose of rewarding the intellectual function that led to the creation of that creative product. It is because the State treats creative products either as an objects of art enriching human culture or as objects of innovation advancing human development, and hence, encouraging their production necessary to secure and maximize public welfare.[22]

The degree of human ingenuity exercised to arrive at a creative product decides the specific modalities of intellectual property rewards that the State ought to grant for that product. When the threshold for human ingenuity in a product to qualify as creative is lower, the strength of enforceability of intellectual property rewards granted by the State over that product is typically also lower, and vice-versa.[23] When the human ingenuity contributed in producing a creative product is shared between two or more individuals, all such persons receive these rewards typically in proportion to their respective contributions.[24]

---

[22] E.g., copyright regimes reward original creations to promote art and enrich human culture, and patent regimes reward novel, non-obvious, and useful creations to promote innovation and advance human development.

[23] E.g., patentable creations demonstrate higher human ingenuity, and are afforded stronger protection against infringement; copyrightable creations demonstrate lesser human ingenuity and are afforded weaker protection against infringement.

[24] E.g., through joint inventorship for patentable creations, and co-authorship for copyrightable creations.

## 2.2 Moral Interest in Creative Products

Like all types of human activity, creativity seeks rewards that could vary in form and magnitude depending on the context in which creativity occurs. But, an essential context-independent reward for creativity is the creator's desire to be known for her creation. This desire to be known is the creator's moral interest – manifested in her creation. The State seeks to protect this interest against abuse by mandating that a creative product must be ascribed to only the creator whose human ingenuity led to its creation, and nobody else.

The State does it as a moral imperative to bring individual justice to the creator (Christie 2011: 10) by granting her what is arguably the foremost reward in an intellectual property regime – *honest ascription*. Honest ascription enables the State to save creators from getting discredited for their creativity so that they do not get discouraged from expressing it. In creative industries, honest ascription also enables a creator to earn a reputation for creativity that is commercially exploitable (on an indefinite basis) so that h/she is constantly encouraged to come up with more creative products (Government of United Kingdom 2006: 51[25]).

## 2.3 Legal Capacity of Creators

For intellectual property rewards to serve an intelligible legal purpose, the person who receives them must demonstrate the necessary legal capacity to: (a) meaningfully enjoy those rewards, and (b) respond to liability claims resulting from harm caused by the nature and use of creative products. (Okediji 2018: 19). Hence, besides serving a moral purpose, honest ascription also serves a functional (legal) purpose.

Sometimes, creative products, by their very nature, might risk harming public order and safety, thereby compromising public welfare. Honest ascription for a creative product enables the State to hold the *true* creator of that product liable for harm and effectively discourage the production of creative products that are *intrinsically harmful*. For instance, Article 19(1)(2) of the Indian constitution imposes reasonable restrictions on the freedom of speech and expression (i.e., copyrightable creations) if it risks harming public order and safety; defaulters (transgressing these restrictions) are prosecuted by the Indian State according to applicable laws. Similarly, inventors are expected to invent responsibly so that their inventions do not pose imminent risks to public order and safety (Stilgoe et al. 2013[26]).

---

[25] Claiming that creators can exploit their celebrity status to make money by appearing in advertising campaigns and by selling branded merchandise.

[26] Proposing that research and innovation must accommodate social and ethical concerns.

Sometimes, creative products, through their misuse, might risk causing harm to public order and safety, thereby compromising public welfare. In all such cases, the State may choose to penalize the owner of the product who is deemed to exercise control over its use. For instance, book publishers could be held liable for irresponsibly disseminating writings that risk harming public order and safety (Seeber and Balkwill 2007: 29). Similarly, product liability laws penalize the manufacturer for non-compliance with due diligence requirements while introducing innovative products (i.e., patentable creations) into the market, risking harm to public order and safety (European Council Directive 85/374/EEC, 25 July 1985).

But, in order to do so, the State must ascertain the legal capacity of the person (natural or artificial) who holds ascription/ownership over that creative product. Only if such a person is legally eligible to anticipate the harmful consequences resulting from the nature or misuse of her creative product, could h/she be held liable for those consequences. This premise holds true for the same jurisprudential reasons that uphold juvenile justice and differential treatment of criminal offences committed by persons of unsound mind.

Hence, minors experience difficulties in enforcing intellectual property rewards even as they are entitled to hold them as creators (Lukose 2013[27]). At the same time, corporates can acquire and enforce intellectual property rewards even as they are not the creators[28]. Corporates are artificial persons that demonstrate legal personhood, i.e., the juridical capability to enter into intellectual property contracts, understand the nature of their contractual obligations, and compensate for harms resulting from their omission or improper performance of those obligations. The State allows corporates to acquire and enforce intellectual property rewards to encourage corporate investment in the production of creative products with the ultimate goal of maximizing public welfare.

## 2.4 Economic Investment in Creative Products

Coming up with creative products could be expensive, demanding significant economic investments. As such, persons (natural or artificial) who invest in creative products would naturally expect good returns on their investments in order to be able to continue with their creative journey.

Ownership of creative products can make that happen. Persons investing in creative products could exercise possession (a key constituent right in ownership) of their products to sell them or rent out their specific uses. However, possession of creative products is almost oxymoronic. This is because creative products are intangible and non-excludable by nature (unlike physical property like real estate and automobiles). Hence, possession of property in the classical sense of physical property cannot be applied to creative products. So, the State grants a set of exclusive rights to the creator over her creative product. These rights constitute the subject matter of ownership for the creator and make possession of the creative product by her creator possible (Christie 2011: 7).

---

[27] Explaining how contractual incapacity disables minors from commercially exploiting their intellectual property.

[28] For instance, through employer-employee/work for hire agreements or through a simple purchase.

But, these exclusive rights have an expiry date, i.e., they remain a time-specific monopoly. This is because the State acknowledges that creative products are non-rivalrous and inexhaustible. In other words, use and enjoyment of a creative product by one person does not hamper the use and enjoyment of that product by another person (Christie 2011: 6-7). Hence, the State concludes that creative products (unlike physical property) could be better utilized by its people through sharing so as to maximize public welfare. Accordingly, the State grants a monopoly on the exclusive rights only for such period of time that ought to enable the creator to recover her economic investment in her creative product, i.e., earn *normal* profits from her creation (Olson 2009); after the expiry of this time period, the creation is ought to be made public.

A time-specific monopoly is a special economic award created by the State only to reward that modicum of human ingenuity that produces creative products (Hughes 1988: 23[29]). In other words, in the absence of that modicum of human ingenuity, time-specific monopolies cannot be justified. It is because monopolies are generally bad. Hence, any use of time-specific monopolies by the State must be extremely restrictive (MacCauley 1841). It should then follow that if human ingenuity in a creative product is reduced, the length of the time-specific monopoly granted for that product must also reduce.

The length of the time-specific monopoly to be granted over the ownership of a certain creative product should be determined by the State as a function of the economic investment put into that product (Sherer 1972[30]). However, this does not mean that higher the economic investment, greater would be the length of the time-specific monopoly. Obviously, market demand of the creative product would play another key role. But, the market demand for any creative product is generally hard to predict (Borisova 2018: 114)[31].

---

[29] Claiming that it is uniqueness of an idea that qualifies it for intellectual property protection.

[30] Proposing that the length of a patent monopoly should be flexible and product-specific, deriving its basis from the economic characteristics of the invention it seeks to reward).

[31] Explaining the difficulties in determining the market potential of creative products that -- unlike products intended for mass consumption – may not necessarily respond to public needs or preferences.

Hence, the State should give precedence to economic investment put into a creative product over its market demand for determining the length of the time-specific monopoly – in which the creator is expected to only recover her economic investment. If the demand for that product in the market becomes favourable, the creator of that product enjoys the liberty to raise its renting or selling price (as in a monopoly, consumers are price takers), thereby enabling herself to earn a *supernormal* profit from the rental or sale of her creative product.

## 2.5 Key Takeaways

To recapitulate, intellectual property rewards for creative products are determined by:

(1) The creator's human ingenuity in her creative product;
(2) The creator's moral interest in her creative product;
(3) The legal capacity of the person (natural or artificial) to enjoy intellectual property rewards granted over the creative product and assume liability for mishaps caused by the nature or misuse of the creative product; and
(4) The economic investment of the person (natural or artificial) in the creative product.

However, to know the necessary justifications for ascription, ownership, and time-specific monopolies granted for creative products, we must break down each of these rewards into the *necessary* factors that ought to determine them. Hence, from our discussion in this chapter, we can say:

(a) Ascription is a function of human ingenuity and moral interest;
(b) Ownership is a function of legal capacity and economic investment;
(c) Monopoly is a function of human ingenuity and economic investment.

The table below summarizes these points:

|  | Human Ingenuity | Moral Interest | Legal Capacity | Economic Investment |
|---|---|---|---|---|
| Ascription | ● | ● | . | . |
| Ownership | . | . | ● | ● |
| Time-Specific Monopolies | ● | . | . | ● |

● = Necessary   . = Sufficient

# 3. Nature of Intellectual Property Rewards

Intellectual property rewards derive their origin from one of the fundamental economic principles which holds that people respond to incentives. Incentives tend to mould human behaviour and could provide great utility in implementing policy changes. However, this principle is not *infallible*:

(1) In the short run, while incentives (as extrinsic stimuli) could induce positive behaviour to achieve a desired policy outcome, they risk weakening "intrinsic motivations" in people to behave desirably in the long-run (Gneezy et al. 2011).

(2) The change in behaviour in response to an incentive cannot always be subject to quantification or other objective assessments.

Intellectual property rewards represent nothing but incentives offered by the State to encourage the production of creative products.

Hence, typical criticisms against incentives moulding human behaviour must also hold true for intellectual property rewards encouraging creators and corporates to produce and invest in creative products.

## 3.1 Intellectual Property Rewards Weaken "Intrinsic Motivations" to Create

It cannot be denied that many of the mankind's greatest creations were produced several hundred years before intellectual property regimes had even begun to develop in various parts of the world. Even when file-sharing broke the Internet in early 2000s, the weakening of the copyright regimes did not correspond to a reduction in the growth of copyrightable creations; on the contrary, it corresponded to a sharp increase in the growth of copyrightable creations (Felix and Koleman 2010[32]).

This implies that the creativity is not *uniquely* dependent on intellectual property rewards. By extension, it would mean that humans possess intrinsic motivations to come up with creative products. Since incentives generally tend to destroy intrinsic motivations in people to behave desirably in the long-run, intrinsic motivations of creators to come up with creative products might gradually weaken in response to intellectual property rewards, if not handled correctly by the State (Amabile and Kramer 2012).

---

[32] Claiming that even as file-sharing technology in recent times has weakened copyright protection, copyrightable creations continue to grow.

## 3.2 Quantification of Change in Creativity Induced by Intellectual Property Rewards is Not Possible

Ideally, the incentive and the intended behaviour change associated with that incentive should ideally be measurable in quantifiable terms. For example, when the government increases the tax rate on the per unit sale of cigarettes, the decrease in cigarette consumption can be measured in quantifiable terms so as to analyze the efficacy of government policy in discouraging people from smoking with a fair degree of objectivity.

Intellectual property rewards cannot be expressed as mathematical functions as most factors determining them cannot be quantified. Hence, the precise economic value of all the intellectual property rewards to be granted for a creative product, for the most part, remains immeasurable. Calculating the economic value of that product, on the other hand, looks more achievable as it can be expressed in terms of its price in the market. But again, the price of a creative product would depend on its market demand that remains hard to predict. Besides, a creative product could be invaluable as well.[33]

Hypothetically, even if one somehow manages to express intellectual property rewards and the value of creative products in quantifiable terms, one still cannot measure the precise impact of intellectual property rewards on creativity in the absence of a counterfactual. Hence, one cannot conclude if intellectual property rewards are *must* for incentivizing creativity.

On the contrary, studies have found that intellectual property rewards can have a negative effect on the production of creative products and public welfare (Torrance and Tomlinson 2009[34]) – and public welfare could be higher with no intellectual property rewards for creative products (Pollock 2008[35]).

Besides, the proliferation in combinatorial creativity (Varian 2010) (Youn et al. 2015) demanding less human ingenuity –  seems to be inspired more by human greed than by motivations to enrich human culture and advance human

---

[33] E.g., several artistic, musical, and literary works that are not intended to be sold by their creators - copyrightable creations.

[34] Claiming that the current patent system (a combination of patent and open source protection) produces significantly lower rates of innovation and societal utility than a commons system.

[35] Claiming that the innovator typically has a first-mover advantage as imitation of innovation is costly – implying that the societal costs actually outweigh the societal benefits accrued from intellectual property.

development — and implies the inefficacy of intellectual property rewards in incentivizing what ought to be transformational creativity (Bell 2006[36]).

Nevertheless, other studies have shown that in the absence of intellectual property rewards creative and innovation industries might slow down — and compromise public welfare (Arrow 1962[37]; Park and Ginarte 1997[38]; Duguet and Lelarge 2012[39]; Posner 2012[40]).

## 3.3 Key Takeaways

Intellectual property rewards though are intended to incentivize the production of creative products, they might gradually weaken intrinsic motivations in creators to come up with creative products. Moreover, the net effect of intellectual property rewards on the production of creative products is moot.

Hence, intellectual property rewards must be used with caution, i.e., these rewards must be granted by the State to reward creativity only after thoroughly examining the factors of human ingenuity, moral interest, legal capacity, and economic investment that ought to determine them -— only towards securing and maximizing public welfare, and nothing else (Chopra 2018).

---

[36] Claiming that copyrights have primarily encouraged entertaining works that sell, and patents have encouraged "marginal improvements in mature technologies".

[37] Claiming that competitive markets by themselves cannot provide incentives for individuals and firms to come up with new inventions; those incentives could be provided only by the government).

[38] Showing that strong intellectual property rights were positively correlated with high investments in research and development.

[39] Claiming that patents overall incentivize firms to come up with product innovations.

[40] Claiming that drug innovation is likely to be discouraged if not granted strong patent protection (as the cost of producing novel drug is very high while producing its identical substitute is very low).

# 4. Modes of Intellectual Property Rewards

The fallible nature of intellectual property rewards demands that the respective modes of their adoption by the State for incentivizing creativity must be accompanied by sufficient caution. While determining a specific reward for a creative product, only the *necessary* factors that ought to determine that reward for the product must be examined, i.e.,

(a) Ascription should be treated as a function of human ingenuity and moral interest;
(b) Ownership should be treated as a function of legal capacity and economic investment; and
(c) Time-specific monopoly should be treated as a function of human ingenuity and economic investment (see Chapter 2.5).

In the current intellectual property practice, intellectual property rewards are codified, i.e., they are conceived and decided *ex ante* by legislative bodies and remain fixed (static) within statutorily designated intellectual property classifications like copyrights and patents.

Copyright is a statutorily designated class of intellectual property protecting original artistic, musical, and literary creations in all their types, forms, and uses. Patent is another statutorily designated class of intellectual property like copyright, except that it protects novel, non-obvious, and useful creations.

Each of these statutorily designated classifications of intellectual property are founded on a common presumption. The presumption is that all creative products within a given classification, copyright or patent, *do not vary* by factors of human ingenuity, moral interest, legal capacity, and economic investment. This presumption basically justifies the possibility of determining intellectual property rewards ex ante for copyrightable and patentable creations.

## 4.1 Ex Ante Determination of Ascription and Ownership is Essential and Justified

Defining rules and regulations for the *ex ante* determination of ascription and ownership for creative products is essential for securing *legal certainty* for creators in the intellectual property regimes.

It is also justified as it is fairly possible to define clear-cut legal standards to represent factors that would be necessary for a claim of ascription (namely, human ingenuity and moral interest) and ownership (namely, legal capacity and economic investment) even before new creative products come into actual existence. In other words, human ingenuity, moral interest, legal capacity, and economic investment could be treated as constants for determining ascription and ownership for creative products.

For example, human ingenuity and moral interest is represented by authorship/co-authorship for copyrightable creations; for patentable creations, they are represented by inventorship/joint inventorship. Ownership of both copyrightable and patentable creations can be assumed and exercised by persons (natural or artificial) who have either economically invested in these creations either by hiring or employing creators or through a simple purchase.

## 4.2 Ex Ante Determination of the Length of Time-Specific Monopolies is Unjustified

Time-specific monopolies over a creative product are intended to enable the creator of that product to only recover her economic investment that were put into the creation. These economic investments actually vary significantly for different creative products – even when these products fall within one designated intellectual property classification, i.e., not all copyrightable creations would bear the same economic investment, and not all patentable creations would bear the same economic investment. Hence, economic investment as a factor for determining the length of the time-specific monopoly over the ownership of a creative product should not be treated as a constant (Thurow 1997[41]; Government of United Kingdom 2006: 39[42]).

Note that here, we are distinguishing between the *existence* of economic investment and the *quantum* of economic investment. Existence of economic investment could be safely treated as a constant for determining ownership of a creative product *ex ante*. However, quantum of economic investment tends to become variable across creative products.

Hence, ex ante determination of a fixed (static) length of time-specific monopolies over the ownership of creative products that remain actually dependent on a variable factor (quantum of economic investment) risks over-rewarding creativity (unduly preventing others from creating – having a chilling effect on creativity) or under-rewarding creativity (failing to sufficiently encourage creativity) (Fraser 2016). This ultimately compromises public welfare.

---

[41] Questioning the rationality of granting patent monopolies of a uniform length (20 years) even as costs across patentable creations would vary significantly.

[42] Claiming that given the extensive range of copyrightable creations, it is not possible to arrive at one "definitive optimal length" for all copyrightable creations.

It also implies that the optimal length of time-specific monopoly to be granted over a creative product (i.e., one that balances the creator's returns on her economic investment in her creative product with public welfare) cannot be determined ex ante. For this reason, many have criticized the long, uniform terms of copyright and patent monopolies (Landes and Posner 2003[43]; Boldrin and Levine 2005[44]; Lester and Zhu 2019[45]).

It should then necessarily follow that public welfare would be better secured through an ex post determination of the length of time-specific monopolies (i.e., after the quantum of economic investment has been ascertained).

## 4.3 Key Takeaways

Statutorily determining ascription and ownership for creative products ex ante is recommended. But, determining the length of time-specific monopolies over the ownership of creative products ex ante grossly discounts the uniqueness and variability of measures of economic investments that they bear – overall risking abuse of time-specific monopolies through over-rewarding creativity (unduly preventing others from creating – having a chilling effect on creativity) or under-rewarding creativity (failing to sufficiently encourage creativity) – ultimately compromising public welfare.

If deciding the time-specific monopoly component of diminished intellectual property rewards for each creative product ex ante looks too wieldy for copyright and patent offices, sub-classifications under copyrightable creations and patentable creations could be devised, where each sub-classification represents and carries products bearing similar quanta of economic investment and market demand.

---

[43] Suggesting that extend the time-length of any intellectual property monopoly beyond 25 years would create little extra incentives for creators.

[44] Suggesting that the optimal length of a copyright monopoly should be seven years.

[45] Quoting economists' criticisms against the practice of granting flat 20-years long patent monopolies on inventions without any regard to their economic characteristics.

# 5. Determinants of Intellectual Property in Artificial Creations

Developers of artificial creators and their proponents claim that an artificial creation is a creative product that is generated by a highly advanced artificially intelligent system autonomously. In this chapter, we check for human ingenuity, moral interest, legal capacity, and economic interest in artificial creations for examining the existence of intrinsic justifications for intellectual property rewards for artificial creativity.

## 5.1 Human Ingenuity in Artificial Creations

Developers of artificial creators and their proponents claim that the exact nature and composition of artificial creations remains unpredictable. Hence, they argue that artificial creators demonstrate a form of "mind-ness" that could replace human creativity in producing creative products. However, the argument is tenable only for the "moment of creation" – i.e., when the artificial creation comes into existence.

    The argument does not hold good for what precedes the moment of creation. The mind-ness exhibited by artificial creators does not come out of thin air; it is meticulously constructed by its developers who tirelessly employ their *human* minds to build it to serve a defined purpose (irrespective of how *broad* that purpose might be). In other words, the mind-ness exhibited by artificial creators has its origins in the minds of their creators, and at least the broad contours of what artificial creators can and cannot do are consciously determined by their developers. Hence, the source of artificial creativity must lie in the *human ingenuity* of its developers (Plotkin 2009: 57-58[46]).

    Simply put, artificial creativity is an artificial function *created* by human ingenuity. The mere fact that the artificial function is unpredictable owing to its broadly defined purpose does not make the artificial creator an autonomous entity like humans. The initial criticality of human ingenuity in producing artificial creations must not be completely discounted (Blok 2017[47]; Shemtov 2019[48]). That

---

[46] Arguing that human ingenuity is not expendable in the invention process *yet*.

[47] Claiming that artificial creators producing patentable creations are mere tools in the inventive process in which human input remains inevitable.

[48] Arguing that conception of the artificial invention should be attributed to the natural person who set up the artificial creator; whether such person could predict the exact nature and composition of the artificial invention should be immaterial to deciding the attribution of its conception).

said, at same time, the absence of human ingenuity (or creativity) in the moment of creation must not be ignored (Abbott 2016a: 1103; Abbott 2016b: 14)[49].

In other words, even as human ingenuity is initially critical to producing artificial creations, its eventual contribution is negligible in producing artificial creations, i.e., the total contribution of human ingenuity in producing artificial creations has *diminished*. A product is deemed creative in intellectual property regimes when it demonstrates at least a modicum of human ingenuity. Hence, it follows that artificial creations should be deemed less creative in comparison to creative products that result from the exercise of pure human ingenuity. Moreover, this diminished human ingenuity put into producing an artificial creation may be shared between two or more natural persons (like for any creative product).

Hence, the State ought to conceive intellectual property rewards that aptly respond to the diminished human ingenuity in an artificial creation and distribute rewards amongst the developers of the artificial creator in proportion to their respective contributions to that diminished human ingenuity.

## 5.2 Moral Interest in Artificial Creations

As stated previously, human ingenuity in artificial creations must not be completely discounted. Hence, honest ascription for an artificial creation would demand that the natural person whose human ingenuity is the source of that artificial creation is not entirely discredited for it as that person must reasonably have at least some moral interest in that artificial creation. At the same time, artificial creators cannot intelligibly demonstrate moral interest in their artificial creations (Perry and Margoni 2010: 627). Hence, the State ought to conceive intellectual property rewards that aptly respond to the reduction in moral interest of the developers of artificial creators in artificial creations.

## 5.3 Legal Capacity of Artificial Creators

Artificial creators cannot foreseeably demonstrate legal capacity either to (a) meaningfully enjoy intellectual property rewards for artificial creations (Perry and Margoni 2010: 627) (Okediji 2018: 19), or (b) respond to liability claims resulting from the harm caused by their nature or misuse of artificial creations. That said, intellectual property rewards that may be granted for artificial creations could be legally enjoyed by the developers of artificial creators whose initial exercise of human ingenuity brought them into existence and/or by artificial persons (corporates) who have economically invested in them. Besides, artificial creations, by their nature or misuse, may risk causing harm to public order and safety. In all such cases, the State ought to fix appropriate liability on persons (natural or artificial) involved in the production or use of artificial creations that led to the harm (Cauffman 2018: 530--32).

---

[49] Arguing that the natural person who set up the artificial creator did not contribute significantly to the patentable creations produced by the artificial creator; hence, recognizing such natural person as the inventor of the artificial inventions would be inefficient and unfair).

## 5.4 Economic Investment in Artificial Creations

Coming up with artificial creators to produce artificial creations is an exorbitant enterprise (Plotkin 2009: 130). As such, persons (natural or artificial) who economically invest in artificial creators and artificial creations would normally expect good returns on their investments in order to be able to continue with artificial creativity. Hence, to protect artificial creativity against economic losses and encourage it, the State ought to grant appropriate intellectual property rewards.

## 5.5 Key Takeaways

Hence, intrinsic justifications for intellectual property rewards for artificial creativity do exist. However, intellectual property rewards need to be adjusted for artificial creativity. General modalities of intellectual property rewards for artificial creations are developed in the next chapter.

# 6. General Modalities for Intellectual Property Rewards for Artificial Creations

In this chapter, we develop general modalities for intellectual property rewards for artificial creations.

## 6.1 Ascription for Artificial Creations

Ascription is a function of human ingenuity and moral interest.

The total contribution of human ingenuity in producing artificial creations gets diminished, implying that they are less creative in comparison to creations that result from the exercise of pure human ingenuity (Burt and Davies 2018: 250)[50] Even so, the source of an artificial creation lies with the human ingenuity of the developer of the artificial creator who must have at least some moral interest in the artificial creation. Hence, honest ascription demands that such person must receive *at least partial credit* for the artificial creation; no ascription for artificial creations risks discouraging human creativity (Fraser 2016[51]) and erasure of public support for intellectual property (Dutfield 2013: 33).

The remaining part of the credit for the artificial creation cannot be granted to the artificial creator as it cannot intelligibly demonstrate moral interest in its creation.

Granting part-ascription for an artificial creation to the natural person whose human ingenuity is the source of that artificial creation serves not just a moral purpose, but also a functional (legal) purpose. Sometimes, artificial creations, by their very nature, might risk causing harm to public order and safety. In all such cases, granting part-ascription for an artificial creation to the person whose human ingenuity is the source of that creation would enable the State to hold that person liable for harm so as to generally discourage artificial creativity that is intrinsically harmful, at least to some extent regardless of the unpredictability of artificial creativity (Stilgoe 2018[52]; Firth-Butterfield and Chae 2018[53]).

---

[50] Arguing that the creator of an AI program that artificially produces a patentable invention autonomously could might at best have a claim of joint inventorship.

[51] Claiming that inventorship enables scientists and engineers to gain professional credibility and monetary benefits from their inventions.

[52] Claiming warning about accepting the fallacious autonomy of self-driving cars that companies might use to shirk liability for mishaps caused by their self-driving cars.

[53] Stressing the need for a degree of human responsibility that ought to give proper direction to innovation practices, in the absence of which innovations would risk leading to unintended and negative consequences.

For instance, an artificial creator might encroach upon the intellectual property of others while producing its artificial creation. In such cases, the developer of that artificial creator should be able to demonstrate due diligence in developing and deploying her artificial creator in order to defend her liability for encroachment upon the intellectual property of others. Else, the developer must be held liable.

But, the developer of the artificial creator must be only part-liable for harm caused by the nature of the artificial creation. It is so because the eventual contribution of the developer of the artificial creation in producing artificial creations is negligible – implying that the exact nature and composition of the artificial creation could not have been predicted by the developer of the artificial creator. On the other hand, imposition of full liability for harm caused by the nature of the artificial creation on the developer of the artificial creator would be unfair and disproportionate and might have a chilling effect on artificial creativity. However, this does not imply that the remaining part of the liability should then be imposed on the artificial creator; the artificial creator cannot foreseeably demonstrate legal capacity to respond to liability claims.

In summary, natural persons whose human ingenuity is the source of artificial creations must be granted part-ascription for those artificial creations so that (a) their moral interests in those artificial creations are protected against abuse, and (b) they remain part-liable for mishaps owing to the nature of those artificial creations to discourage artificial creativity that is intrinsically harmful.

## 6.2 Ownership of Artificial Creations

Ownership is a function of legal capacity and economic investment (existence).

Producing artificial creations could be an exorbitant enterprise (Plotkin 2009: 130). But, artificial creations cannot be owned like physical property to derive economic gains by selling or renting it. Artificial creations (like other creative products) are intangible and non-excludable by nature, and therefore, their physical possession is impossible. Hence, the State ought to grant a set of exclusive rights to the use and enjoyment of artificial creations for a limited period of time. This set of exclusive rights to the use and enjoyment of artificial creations constitutes the subject matter of ownership of artificial creations to make the possession of artificial creations possible for their natural creators or artificial persons who have economically invested in them.

The owner of an artificial creation may either be the developer of the artificial creator whose initial exercise of human ingenuity brought the artificial creation into existence (by virtue of receiving part-ascription for the artificial creation), or an artificial person (corporate) who acquires ownership from that developer (through employer-employee/work for hire agreements or through a simple purchase). The owner may choose to personally use or enjoy the artificial creation or sell it or rent out its certain uses to derive economic gains. The owner is also legally responsible for harms caused to public order and safety by the misuse of her artificial creation (as h/she could be reasonably presumed to exercise control over the use of her creation). Hence, the owner of the artificial creation must respond to liability claims resulting from harms to public order and safety caused by the misuse of her artificial creation.

To meaningfully acquire and exercise ownership over artificial creations and to respond to liability claims resulting from harm caused by the misuse of artificial creations, the owner must demonstrate legal capacity. Hence, the owner of an artificial creation could be the natural person in whose human ingenuity lies the source of the creation, or an artificial person (corporate) who may has economically invested in the creation – as these persons are legal persons. Artificial creators, on the other hand, do not demonstrate the legal capacity to meaningfully acquire and enforce ownership over their artificial creations or respond to liability claims resulting from harm caused by the misuse of their artificial creations (Abbott 2016a: 1107)[54]

Yet, there might be another concern around the ownership of artificial creations. Ownership of a creative product in intellectual property regimes flows from ascription over that product (i.e., creators are deemed first owners except if otherwise has been pre-determined by employer-employee/work for hire agreements). One might argue that part-ascription for an artificial creation justifies only its part-ownership. Part-ownership over an artificial creation might translate into reducing the number of exclusive rights to be granted over that artificial creation or the strength of enforceability of those rights or both.

But, then it must be balanced with the imposition of part-liability on owners for harms caused to public order and safety caused by the misuse of artificial creations. However, reduction in the liability of owners of artificial creations might seriously jeopardize public order and safety. If, on the other hand, liability of owners of artificial creations is not reduced in response to reduction in ownership (in the form of part-ownership), artificial creativity might end up being discouraged (Viscusi and Moore 1993[55]).

---

[54] Arguing that artificial creators only act upon the instructions of its natural creator and remain entirely agnostic about the use of their creations.

[55] Arguing that "misdirected liability may depress beneficial innovations".

In summary, full ownership of artificial creations should be granted either to the developer of the artificial creator that produced that artificial creation (by virtue of receiving part-ascription for the artificial creation), or an artificial person (corporate) who acquires ownership from that developer (through employer-employee/work for hire agreements or through a simple purchase) to hold them fully liable for the misuse of their artificial creations – towards securing public order and safety.

## 6.3 Time-Specific Monopolies over Ownership of Artificial Creations

Time-specific monopoly is a function of human ingenuity and economic investment (quantum).

If human ingenuity in a creative product is reduced, the length of the time-specific monopoly granted for that product must also reduce (see Chapter 2.4).

Human ingenuity in artificial creations gets diminished. Hence, the length of the time-specific monopoly granted for an artificial creation must also reduce. The reduced time-period of such a monopoly must be less than what ought to be granted for a creative product that is result of the exercise of pure human ingenuity. If the length of the time-specific monopoly is not reduced in this way, human ingenuity (arguably, the source of many more artificial creators to come) would get undermined. Note that this is assuming the respective economic investments for the creative product and the artificial creation to be equal.

At the same time, the length of the time-specific monopoly granted for an artificial creation should also be a function of the quantum of economic investment that was put into that artificial creation. That is, the length of the such a monopoly should be just enough to enable the developer of the artificial creator which produced that artificial creation to recover her economic investment, i.e. earn normal profits from selling the artificial creation or renting out its certain uses. As for any other creative product, if the market demand for the artificial creation becomes favourable, such a monopoly would likely enable such developer of the artificial creator to increase the selling or renting price of her artificial creation (as in a monopoly, the consumers are price takers), and earn supernormal profits from selling the artificial creation or renting out its certain uses.

Such time-specific monopolies over the ownership of artificial creations could only be granted ex post as economic investments for them (even within the same statutorily designated intellectual property classification) might vary significantly. If, however, they are granted ex ante, i.e., without considering the quanta of economic investments that artificial creations may bear, they would risk under-rewarding artificial creations (discouraging artificial creativity) and over-rewarding artificial creations (undermining human ingenuity).

## 6.4 Key Takeaways

From the discussions in this chapter, we can conclude that intellectual property rewards for an artificial creation should include: (a) part-ascription for the developer of the artificial creator whose initial exercise of human ingenuity brought that artificial creation into existence, and (b) reduced length of the time-specific monopoly over the ownership of that artificial creation (in comparison to that of a creative product resulting from pure human ingenuity bearing equal economic investment) to be granted either to the developer of the artificial creator or the artificial person (corporate) who has economically invested in the artificial creation. The nature and composition of the set of exclusive rights (i.e., subject matter of ownership) should remain the same for artificial creations (falling under the same statutorily designated classification of intellectual property – copyright or patent).

Hence, the sum of rewards granted for artificial creations to incentivize artificial creativity has overall diminished in comparison to that for creative products resulting from the exercise of pure human ingenuity. This diminished sum of rewards for artificial creations is *diminished intellectual property*. In other words, intellectual property rewards for artificial creations are *diminished intellectual property rewards*.

The table below summarizes these points:

|  | Human Ingenuity | Moral Interest | Legal Capacity | Economic Investment |
|---|---|---|---|---|
| Part-Ascription | ◔ | ◔ | • | • |
| Ownership | • | • | ● | ● |
| Reduced Time-Specific Monopolies | ◔ | • | • | ● |

= Necessary (Diminished) Factor;  **.** = Sufficient Factor

# 7. Critique of the Treatment of Artificial Creativity by Current Intellectual Property Regimes

This chapter contains a critique of the treatment of artificial creativity by current intellectual property regimes in different jurisdictions based on the intrinsic justifications for intellectual property rewards for artificial creativity and the general modalities for these rewards developed in the previous chapter.

Laws protecting copyrightable and patentable creations remain typically distinct from each other, and therefore, their treatment of artificial creativity have been discussed under separate heads depending on whether the artificial creation is copyrightable ("artificial work") or patentable ("artificial invention").

## 7.1 Artificial Works

Artificial works are creative products that would be typically copyrightable under copyright regimes if created by pure human ingenuity (i.e., by a human without using an artificial creator).

**United Kingdom.** Laws in the United Kingdom do grant intellectual property rewards for artificial works. Section 178 of the UK Copyright Designs and Patents Act, 1988 ("CDPA 1988") recognizes artificial works as works in which human authorship is absent.[56] The provision is plausible as human authorship should demand full ascription for those artificial works to the natural person who developed the artificial creator. But, such natural person deserves only part-ascription for those artificial works as her human ingenuity would be negligible in the moment of creation of those artificial works (as discussed in Chapter 5.1).

Section 9(3) of the CDPA 1988, however, is problematic for this very reason. It assigns authorship or full ascription for the artificial work to the developer of the artificial creator (Burt and Davies 2018: 244)[57].

It is a disproportionate reward of ascription which risks undermining human ingenuity (arguably, the source of many more artificial creators to come) and might also unfairly invite legal liability for the person if the artificial work by its nature harms public order and safety – creating a chilling effect on artificial creativity (as discussed in the Chapter 6.1).

---

[56] E.g., Nova Productions Ltd. v. Mazooma Games Ltd. & Ors. Rev. 1 [2006] EWHC 24 (Ch) at 104.

[57] Arguing that such an assignment is a creation of legal fiction without any tenable justification.

The subject matter of ownership that is granted for artificial works under CDPA 1988 bears a semblance of neighbouring rights in copyright (Bently 2018). This is problematic for reasons discussed in Chapter 6.2.

The time-specific monopoly that is granted over the ownership of artificial works under CDPA 1988 is reduced to 50 years. It might be justified to the extent that another work being a product of pure human ingenuity and bearing equal economic investment receives a time-specific monopoly grant which is at least more than 50 years. However, the British Parliament has decided the 50-year period ex ante, i.e., by refusing to decide the length of the time-specific monopoly as also a factor of economic investment (quantum) which could vary for different artificial works that are yet to be generated by artificial creators.

***Europe.*** European laws and jurists do not endorse intellectual property rewards for artificial works. Even as European laws do not expressly prohibit granting intellectual property rewards for artificial works, European jurists argue that artificial works are not original. They do so by citing European judicial opinions holding that for a work to be original, it must reflect the personality of the author – implying that artificial works do not reflect the personality of the author (Iglesias et al. 2019: 14).

This argument is problematic. It is plausible to reckon that artificial works – for want of human authorship – may carry less of the personality traces of the natural person who developed the artificial creator. Even so, these traces are bound to exist (Shemtov 2018). They may or may not be clearly identifiable by everyone at all times. Nevertheless, they could always be argued to exist – at least in part – implying that the work generated by an artificial creator is original – at least in part – and entitling the natural person who developed the artificial creator to diminished intellectual property rewards.

The European position in holding that artificial works are not original preempts diminished intellectual property rewards for artificial works to the natural person who developed the artificial creator. Hence, according to the European position, artificial works ought to fall into the public domain – a proposal by some legal scholars (Perry and Margoni 2010; Schönberger 2018).

However, this is likely to discourage the investment in artificial creativity (Iglesias et al. 2019: 15). Lately, businesses have started using artificial works on a proprietary basis to advertise their products and services. Adoption of the European position on artificial works would discourage these businesses from doing so (Guadamuz 2017a, 2017b).

***United States.*** American laws expressly prohibit granting intellectual property rewards for artificial works. The Compendium of US Copyright Office Practices (like section 178 of the UK CDPA 1988) acknowledges the absence of human authorship in artificial works, but unlike section 9(3) of the UK CDPA 1988 does not assign authorship (or full ascription) for artificial works to the developers of artificial creators.

It does not allow any form of intellectual property rewards for artificial works, ultimately leading to similar policy outcomes led by the European position on artificial works.

***India.*** Indian laws (like UK laws) via Section 4(d)(iv) of the Indian Copyright Act, 1957 grant intellectual property rewards for artificial works by assigning authorship (or full ascription) of artificial works to the developer of the artificial creator, leading to similar policy outcomes led by the English position on artificial works.

## 7.2 Artificial Inventions

Artificial inventions are creative products that would be typically patentable under patent regimes if created by pure human ingenuity (by a human without using an artificial creator).

***United Kingdom.*** In the United Kingdom section 7(2) the Patents Act, 1977 defines the inventor as the "actual deviser of the invention". The Act does not expressly mandate that such an inventor must be a natural person (Davies 2018: 248). Hence, in principle, artificial creators producing artificial inventions could be deemed inventors. However, that would (a) neglect the lack of moral interest and legal capacity of artificial creators to meaningfully enjoy inventorship over their artificial inventions (as discussed in Chapter 5.2 and Chapter 5.3), and (b) deny partial credit for the artificial invention that ought to be granted to the developer of the artificial creator, ultimately over-rewarding artificial creativity, undermining human ingenuity, and erasing public support for intellectual property (or creativity) (as discussed in Chapter 6.1).

***Europe.*** Like in the UK, European laws do not expressly provide that the inventor must be a natural person (Iglesias et al. 2019: 16-17) – ultimately leading to similar policy outcomes led by the UK position on artificial inventions.

***United States.*** United States laws differ from the UK and European positions on artificial inventions in that they mandate the inventor to be a natural person.[58] The position of United States laws on artificial inventions would discourage artificial creativity that could produce numerous useful innovations in drug development, product design, etcetera.

***India.*** Indian laws do not expressly mandate that the inventor must be a natural person (Indian Patents Act, 1970; Soni and Singh 2019). Hence, they might lead to the similar policy outcomes led by the English position on artificial inventions.

In the absence of inventorship for artificial inventions, artificial inventions could be unpatentable in most jurisdictions – a proposal by some American legal scholars as well (McLaughlin 2018). This means that artificial inventions shall not be entitled for intellectual property rewards – discouraging artificial creativity that shows tremendous potential in drug development, design innovation, etcetera.

To correct this situation, even in the absence inventorship, *diminished intellectual property rewards* must be granted for an artificial invention to the developer of the artificial creator which produced that artificial invention.

## 7.3 Key Takeaways

From the discussions in this chapter, it is clear that the treatment of artificial creativity by current intellectual property in different jurisdictions is either improper or inadequate. Realizing the humungous business potential offered by artificial works and artificial inventions, the creative and innovation industries have started lobbying for law reforms in the intellectual property regime to accommodate artificial works and artificial inventions (Tata Consultancy Services and Confederation of Indian Industry 2019). Hence, law-makers, jurists, and most importantly, those who create and invest in artificial creators must come together to negotiate intellectual property rules that are proper and adequate to reward artificial creativity. In the next chapter, I propose a blueprint for granting specific intellectual property rewards for artificial works and artificial inventions.

---

[58] 1 U.S.C. Sec. 8(a) r/w 1 U.S.C. Sec. 1; Beech Aircraft Corp. v. EDO Corp., 990 F.2d 1237, 1248 n.23 (Fed. Cir. 1993); Karrer v. United States, 152 F.Supp. 66, 69 (Ct. Cl. 1957).

# 8. Specific Intellectual Property Rewards for Artificial Works and Artificial Inventions

Intellectual property regimes should reward artificial creativity in a way that secures an optimal balance between incentivizing their production and enabling access to their use by creators and the society at large so as to ultimately maximize public welfare. This chapter discusses the two primary statutorily designated classifications of intellectual property, i.e., copyrightable creations and patentable creations, the core jurisprudential criteria defining them and enabling their classification, and how these criteria need to adapt to accommodate artificial works and artificial inventions.

## 8.1 Artificial Works

Copyrights are granted for creations that *inter alia*[59] demonstrate originality – a modicum of human ingenuity (or creativity). Originality only means that the creation must emanate independently from an intellectual mind. That intellectual mind is deemed the author of that creation. Independent creation implies that the creation must not be copied from an already existing creation. A copyrighted creation receives protection against infringement only if the infringing material is the result of copying from that creation, i.e., independent reproduction of a copyrighted creation is a valid defence to a claim of copyright infringement of that creation (Christie 2011: 14).

The jurisprudential criteria for defining copyrightable creations makes it clear that ascription for a copyrightable creation is realized through authorship. Ownership of a copyrightable creation includes exclusive rights of reproduction, adaptation, sale, transfer through rental or lending, performance, and display of that creation by the owner. The time-specific monopoly granted over the ownership of copyrightable creations is typically fixed at the lifetime of the author – plus 50 to 100 years (depending on the jurisdiction).

**Specific Modalities.** If a copyrightable creation is an artificial work, it ought to be protected by a *diminished copyright*. In line with the conclusions made in Chapter 6, diminished copyright rewards for artificial works imply the following:

(1) Part-ascription for the artificial work ought to be granted to the developer of the artificial creator whose initial exercise of human ingenuity led to the artificial work. Part-ascription here does not imply part-authorship or co-authorship. Authorship implies independent creation, i.e., originality, which is absent in artificial works as they could be deemed original only in part; human ingenuity remains absent in the moment of creation for artificial works (Marcus and Goncalves 2019: 72-75). A title of part-ascription for an artificial work granted to the developer of the artificial creator could be called *post-authorship* (i.e., something that is *second* to authorship).

---

[59] To attract copyrightability, creations must also be fixed in tangible media.

Part-ascription for an artificial work could be shared between the developers of an artificial creator in proportion to their initial contributions of human ingenuity that led to the artificial work. All such developers of the artificial creator should be granted the title of *joint post-authorship* over the artificial work.

(2) The subject matter of copyright ownership over artificial works, i.e., the set of exclusive rights to be granted over artificial works ought to remain unchanged (for reasons discussed in Chapter 6.2). The subject matter of ownership could be shared via contractual arrangements between developers of artificial creators and persons (natural or artificial) who economically invested in the artificial works.

(3) The length of the time-specific monopoly to be granted over the artificial work must be determined ex post (i.e., after the work comes into existence). It should be determined as a function of the quantum of the economic investment that the artificial work carries and should be just enough to enable the person (natural or artificial) to recover that investment, i.e., earn normal profits from selling that artificial work or renting out its certain uses. A secondary consideration for determining the length of the time-specific monopoly to be granted over the artificial work could include an assessment of the market demand of the artificial work.

If the market demand for the artificial work becomes favourable, such a monopoly would likely enable the person (natural or artificial) to increase the selling or renting price of the artificial work (as in a monopoly, the consumers are price takers), and earn supernormal profits from selling such sale or rental.

Diminished copyright rewards for artificial works would ensure that artists and businesses are sufficiently incentivized to economically invest in them, use them for their branding, or sell them or rent out their certain uses in the market to earn profits, without reducing incentives for the production of works of pure human ingenuity that make creation of objects of art intrinsically meaningful and truly enrich human culture.

**Concerns.** The implementation of diminished copyrights for artificial works might lead to some unintended policy outcomes in copyright regimes.

For instance, the prospect of diminished intellectual property rewards might lead persons (natural or artificial) to deliberately hide the involvement of artificial creativity in producing artificial works from the copyright office – so that artificial works could be registered with the copyright office as creative products resulting from pure human ingenuity to attract greater copyright rewards, i.e., authorship (or full ascription) and longer time-specific monopolies.

In such situations, copyright offices might risk over-rewarding artificial works because artificial works and works of pure human ingenuity tend to look extremely similar to each other and copyright offices might lack the capabilities to detect differences between the two (Michaux 2018). Such equal treatment of artificial works and works of pure human ingenuity would ultimately undermine and discourage creation of the latter (Schönberger 2018[60]; Lauber-Rönsberg and Hetmank 2019[61]). Hence, requirements for disclosure of involvement of artificial creativity could be made mandatory, and pecuniary penalties could be imposed on defaulters so as to discourage intentional non-disclosure. That said, evidence suggests that businesses willingly disclose the involvement of artificial creativity in artificial works they use – as a novel marketing strategy to sell these artificial works at enormous prices (Cherie 2019).

## 8.2 Artificial Inventions

Patents are granted for useful creations that demonstrate novelty and non-obviousness/inventiveness – a much higher degree of human ingenuity (or creativity) than what copyright grants typically demand. A creation is deemed novel if the publicly available prior art does not mention or disclose it. It is deemed non-obvious/inventive if a person with ordinary skill in the art could not have created it *obviously*. A patented creation receives protection against infringement irrespective of whether the infringing material is the result of copying from that creation, i.e., independent reproduction of a patented creation is not a valid defence to a claim of patent infringement of that creation (Christie 2011: 14).

Ascription for a patentable creation is realized through inventorship. An inventor is a person who conceives the invention and reduces it to practice. Ownership of a patentable creation includes exclusive rights of use, manufacture, and transfer through sale or rental of that creation by the owner. The time-specific monopoly granted over the ownership of patentable creations is typically fixed at 20 years across jurisdictions.

---

[60] ibid (n14).

[61] ibid (n16).

***Specific Modalities.*** If a patentable creation is an artificial invention, it ought to be protected by a *diminished patent*. In line with the findings in Chapter 6, diminished patent rewards for artificial works imply the following:

(2) Part-ascription for the artificial invention ought to be granted to the developer of the artificial creator whose initial exercise of human ingenuity brought the artificial invention into existence. Part-ascription here does not imply part-inventorship or joint inventorship. Inventorship implies *conception* of an invention and its reduction into practice,[62] which is absent in artificial inventions as their nature and composition remain unpredictable to the developer who created the artificial creator; human ingenuity remains absent in the moment of creation of the artificial inventions. A title of part-ascription for an artificial invention granted to the developer of the artificial creator could be called *post-inventorship* (i.e., something that is *second* to inventorship).

Part-ascription for an artificial invention could be shared between developers of the artificial creator in proportion to their initial contributions of human ingenuity that led to the artificial invention. All such developers of the artificial creator should be granted the title of *joint post-inventorship* over the artificial invention.

(2) The subject matter of patent ownership over artificial inventions, i.e., the set of exclusive rights to be granted over artificial inventions ought to remain unchanged (for reasons discussed in Chapter 6.2). The subject matter of ownership could be shared via contractual arrangements between developers of artificial creators and persons (natural or artificial) who have economically invested in the artificial inventions.

(3) The length of the time-specific monopoly to be granted over ownership of the artificial invention must be determined ex post (i.e., after the invention comes into existence). It should be determined as a function of the quantum of the economic investment that the artificial invention carries and should be just enough to enable the person (natural or artificial) to recover that investment, i.e., earn normal profits from selling that artificial invention or renting out its certain uses. A secondary consideration for determining the length of the time-specific monopoly to be granted over the artificial invention could include an assessment of the market demand of the work.

If the market demand for the artificial invention becomes favourable, such a monopoly would likely enable the person (natural or artificial) to increase the selling or renting price of the artificial invention (as in a monopoly, the consumers are price takers), and earn supernormal profits from selling such sale or rental.

---

[62] Townsend v. Smith, 36 F.2d 292 (1929).

Diminished patent rewards for artificial inventions would ensure that inventors and businesses are sufficiently incentivized to economically invest in them, use them for their branding, or sell them or rent out their certain uses in the market to earn profits – without reducing incentives for the production of inventions of pure human ingenuity (i.e., inventions conceived and reduced to practice by human inventors) that remains the primary source of the most groundbreaking inventions till date, including artificial creators, and arguably, truly inspires the innovation economy still.

**Concerns.** However, the implementation of diminished patents for artificial inventions might create some friction in the patent regimes –

(a) One might apprehend the absurd possibility that diminished rewards for artificial inventions might constrain businesses to abandon the use of artificial creators and employ only pure human ingenuity instead to produce inventions that could receive greater patent rewards, i.e., inventorship (or full ascription) and longer time-specific monopolies (Abbott 2016a). But, this apprehension is likely to be false.

Artificial creators are typically great at producing artificial inventions that bear a strong semblance of combinatorial innovations. They do so by carrying out millions of simulations in no time and at reduced costs (Vertinsky and Rice 2002: 576; King et al. 2009: 47; Burt and Davies 2018: 249) demonstrating a significant comparative advantage to human inventors (Knorr 2001; Marks 2015: 10, 32-35)[63] in producing combinatorial innovations.

Hence, the *seeming undesirability* of diminished patent rewards for artificial inventions should be set off by the ability of businesses to produce useful combinatorial innovations with greater efficiency if they use artificial creators – enabling them to outsell their counterparts operating without artificial creators (i.e., human inventors). Hence, while combinatorial innovations get relegated to artificial creators, pure human ingenuity (i.e., human inventors operating without artificial creators) will only get pushed to come up with inventions that are not merely combinatorial.

(b) There have been efforts to artificially conceive and publish several possible prior art to eventually precipitate the unpatentability of future inventions

---

[63] Arguing that pure human ingenuity is ill-equipped to solve many highly complex problems, and hence, unable to conceive many useful forms of innovation, like those dealing with the multiplicity of variables in nanotechnology and biotechnology.

– both artificial and those produced by pure human ingenuity.[64] Some might view it as unfairly discounting the novelty, and thereby patentability, of future human inventions, ultimately discouraging it. However, this view underestimates the potential of pure human ingenuity. Publication of prior art conceived artificially will only end up raising the bar for novelty that will only push human ingenuity to come up with truly novel inventions (with or without artificial creators) with great potentials to benefit the innovation economy. Moreover, it would also enable the State to exclude frivolous artificial inventions (cheaply produced with little practical use) from the scope of patentability (Vertinsky and Rice 2002: 608).

(c) Uncertainties lie around deciding the person having ordinary skill in the art (or PHOSITA) for the determination of non-obviousness of artificial inventions (Firth-Butterfield and Chae 2018). In 2015, a UK court in Teva v. Leo Pharma[65] explained that an invention was non-obvious so long as there was "no reasonably optimistic expectation" for its conception. Pertinently, the ruling was made in the context of an artificial invention. Hence, determining PHOSITA to determine the non-obviousness of an artificial invention, it is essential that a reasonably optimistic expectation of an invention from that PHOSITA could be derived.

A reasonably optimistic expectation of an artificial invention cannot be derived for artificial creators so long as their internal processes remain unpredictable (Blok 2017). Hence, proposals for raising the non-obviousness standard made out to be necessary – assuming the predominant use of certain artificial creators to invent products and processes in a certain industry (Abbott 2019[66]) – are bound to remain vague, imprecise, incomprehensible, and impracticable.

On the other hand, predictability of artificial creators would imply that the *conception* of artificial inventions must be attributed to the developers of artificial creators. Hence, a reasonably optimistic expectation of an artificial invention could be derived only for the developers of artificial creators. In other words, human inventors operating without artificial creators should remain the PHOSITA to determine the non-obviousness of an artificial invention.

Some apprehend that doing so might put human inventors operating without artificial creators at a disadvantage in relation to their counterparts operating with artificial creators (Fraser 2016; Abbott 2016a). However, again, that disadvantage would be set off by the ability of these human inventors (operating without artificial creators) to claim longer patent monopolies over their inventions in comparison to diminished patent monopolies over artificial inventions granted to their counterparts (operating with artificial creators).

---

[64] *See* 'All Prior Art' project by Alexander Reben, (https://areben.com/project/all-prior-art/) (accessed on 10 April 2020).

[65] [2015] EWCA Civ 588.

[66] Claiming that artificial creativity would eventually render all forthcoming innovations obvious, and thereby unpatentable under the current patenting norms.

(d) Traditional patent law demands an enabling disclosure from the inventor for the invention on which a patent is requested. This enabling disclosure must enable a PHOSITA to reduce the invention to practice (i.e., reproduce it) without employing any additional skill or experimentation. This demand of enabling disclosure for an invention by the State seeks to secure access to the invention by (i) fellow inventors who could build other useful inventions using the enabling disclosure, and (ii) businesses who can scale the production of the invention (after the expiry of the patent) and allow more consumers to benefit from it with cheaper prices. In this way, the State ensures that incentivizing inventions through patents is not carried out by compromising public welfare.

Artificial creations remain *inexplicable* to developers of artificial creators which produce these artificial creations. Hence, artificial inventions remain unpatentable under traditional patent law since a complete enabling disclosure akin to that for a human invention cannot be provided for them (World Intellectual Property Organization 2019b). However, diminished patents for artificial inventions may not need to comply with the expectations of traditional patent law. Diminished patent rewards could be argued to justify a partial relaxation of enabling disclosure requirements for artificial inventions. The requirements for the resulting disclosure for an artificial invention could be just enough to keep artificial creativity from becoming a trade secret. These requirements could be steadily developed by patent offices by analyzing the data on patent applications for artificial creations (Wamsley 2011).

## 8.3 Key Takeaways

From our discussions in this chapter we can summarize that diminished intellectual property rewards for artificial creations should comprise of:

(1) Part-ascription over the artificial creation for the developer of the artificial creator; titles of authorship/co-authorship for copyrightable creations and inventorship/joint-inventorship for patentable creations found in traditional intellectual property law ought to be discarded and replaced with titles of post-authorship/joint post-authorship for artificial works and post-inventorship/joint post-inventorship for artificial inventions to be granted to the developer(s) of the artificial creator;

(2) Full ownership of the artificial creation for the developer of artificial creator and/or persons (natural or artificial) who have economically invested in the artificial creation;

(3) Time-specific monopoly over the ownership of the artificial creation to be decided ex post based on the quantum of economic investment that the artificial creation bears and its market demand. If deciding the time-specific monopoly component of diminished intellectual property rewards for each artificial creation ex ante looks too wieldy for copyright and patent offices, sub-classifications under artificial works and artificial inventions could be devised, where each sub-classification represents and carries artificial creations bearing similar quanta of economic investment and market demand.

Hence, we can conclude that traditional copyright and patent rewards – their modalities and the criteria for their grant – do not need to be completely rejected or replaced with new rewards, but their modalities and the criteria for their grant must evolve to correspond to the uniqueness and variety of artificial works and artificial inventions.

# 9. Conclusion

A product is deemed creative when it demonstrates human ingenuity and exhibits potential to enrich human culture and advance human development. The function of intellectual property is to encourage the human ingenuity of the creator behind the creative product to produce more creative products. This secures the continuous enrichment of human culture and advancement of human development towards securing and maximizing public welfare. Artificial creations vary from creative products in that they demonstrate less human ingenuity (or creativity) in comparison to creative products, even as, arguably, both could hold equivalent potential to enrich human culture and advance human development.

Since artificial creations are not the same as creative products, they demand differential or exceptional treatment by the State. Artificial creations ought to be rewarded by the State through diminished intellectual property (like diminished copyrights for artificial works and diminished patents for artificial inventions). If not, it would lead to over-rewarding artificial creativity and under-rewarding human creativity, ultimately compromising public welfare.

Pertinently, this proposal for diminished intellectual property for artificial creations does not deviate fundamentally from the jurisprudence of intellectual property, but only provides a restructuring of its modalities to recognize and reward the peculiarity of artificial creativity by holding considerations of public welfare paramount. This *restructuring* (as opposed to a fundamental overhaul) of the modalities of intellectual property should be administratively and politically feasible for adoption by intellectual property regimes across the world due to the incremental nature of the policy improvement it proposes to recognize and reward artificial creativity (Lindblom 2016).

If artificial creators are to eventually precipitate the redundancy of human creativity as argued by some (Plotkin 2009: 8; Abbott 2016b: 1; Abbott: 2016b: 13), and artificial creations (instead of human creations) would sought to be encouraged by the State to enrich human culture and advance human development, then intellectual property might not be the right legal/economic tool to achieve it.

# Appendix: Glossary

| Term | Meaning |
|---|---|
| **Artificial Creation** | A creative product that is generated by a highly advanced artificially intelligent system autonomously. |
| **Artificial Creativity** | The phenomenon of artificial creators producing artificial creations. |
| **Artificial Creator** | A highly advanced artificially intelligent system that is capable of producing creative products autonomously. |
| **Artificial Invention** | A creative product that would be typically patentable under patent regimes if created by pure human ingenuity (i.e., by a human without using an artificial creator). |
| **Artificial Work** | A creative product that would be typically copyrightable under copyright regimes if produced if created by pure human ingenuity (i.e., by a human without using an artificial creator). |
| **Ascription** | An intellectual property reward granted by the State for a creative product that is necessarily determined by the contribution of human ingenuity and exhibition of moral interest of the creator in her creative product. |
| **Creative Product/ Human Creation/ Creation** | A product that is created by a natural person and that demonstrates a modicum of human ingenuity and potential to enrich human culture and advance human development. |
| **Creativity/Human Creativity** | The phenomenon of natural persons producing creative products. |
| **Creator/Human Creator** | A natural person who produces a creative product using only her human ingenuity (i.e., by not using an artificial creator). |
| **Economic Investment (Existence)** | The attribute in a creative product that enables ownership of exclusive rights over that creative product by an artificial person like a company. |

| **Economic Investment (Quantum)** | The attribute in a creative product that ought to decide the length of the time-specific monopoly to be granted by the State over the ownership of that creative product. |
|---|---|
| **Human Ingenuity** | A primary qualifier (broadly connoting originality) for a product to be deemed creative. |
| **Intellectual Property** | Intellectual property is granted by the State to the creator for her creative product and includes a set of rewards, namely, ascription – acknowledging the creator, ownership – of a set of exclusive rights over the creation, and time-specific monopolies – setting the expiry date for those exclusive rights. |
| **Legal Capacity** | The juridical capability of a creator to enjoy the intellectual property in her creative product and to respond to liability claims resulting from her creative product or its misuse. |
| **Moral Interest** | The desire of a creator to be known for her creative product that ought to be derived from the contribution of her human ingenuity into her creative product and that enables her to receive ascription over her creative product in proportion to her contribution in her creative product. |
| **Ownership** | An intellectual property reward granted by the State for a creative product that is necessarily determined by the legal capacity of the person (natural or artificial) and existence of economic investment in the creative product by that person to whom it ought to be granted. |
| **Time-Specific Monopoly** | An intellectual property reward granted by the State over the ownership of a creative product that is necessarily determined by the contribution of human ingenuity of the creator and the quantum of economic investment in the creative product by the person (natural or artificial) to whom it ought to be granted. |

# References

Abbott, R. 2019. 'Everything Is Obvious'. *UCLA Law Review* 66 (2), (https://www.uclalawreview.org/everything-is-obvious/) (accessed on 8 March 2020).

Abbott, Ryan. 2016a. 'I Think, Therefore I Invent: Creative Computers and the Future of Patent Law'. *Boston College Law Review* 57 (4), (http://lawdigitalcommons.bc.edu/bclr/vol57/iss4/2) (accessed on 20 February 2020).

Abbott, Ryan. 2016b. 'Hal the Inventor: Big Data and Its Use by Artificial Intelligence' in Sugimoto Cassidy R. et al. (eds.): *Big Data Is Not a Monolith*. Cambridge: MIT Press.

AIVA. 'About AIVA – Artificial Intelligence Visual Artist' (https://www.aiva.ai/about) (accessed 10 March 2020).

Amabile and Kramer. 2012. 'What Doesn't Motivate Creativity Can Kill It'. *Harvard Business Review*, (https://hbr.org/2012/04/balancing-the-four-factors-tha-1) (accessed on 1 March 2020).

Arrow, Kenneth. 1962. 'Economic Welfare and the Allocation of Resources for Invention', in Universities-National Bureau Committee for Economic Research, Committee on Economic Growth of the Social Science Research Council (ed.): *The Rate and Direction of Inventive Activity: Economic and Social Factors*. Princeton University Press, pp. 609–626, (https://www.nber.org/chapters/c2144.pdf) (accessed on 28 March 2020).

Bell, Tom W. 2006. 'Prediction Markets for Promoting the Progress of Science and the Useful Arts'. *George Mason Law Review* 14: 37, (https://ssrn.com/abstract=925989) (accessed on 8 March 2020).

Bently, L. 2018. 'The UK's provisions on computer generated works: a solution for AI creations?'. *ECS International Conference: EU copyright, quo vadis? From the EU copyright package to the challenges of Artificial* Intelligence, (https://europeancopyrightsocietydotorg.files.wordpress.com/2018/06/lionel-the-uk-provisions-on-computer-generated-works.pdf) (accessed on 16 March 2020).

Blok, Peter H. 2017. 'The Inventor's New Tool: Artificial Intelligence – how does it fit in the European Patent System'. *European Intellectual Property Law Review* 39 (2): 69-73, (https://dspace.library.uu.nl/handle/1874/372854) (accessed on 10 March 2020).

Boldrin, M. and David Levine. 2005. 'Growth and Intellectual Property' *National Bureau of Economic Research*, (https://www.nber.org/papers/w12769.pdf) (accessed on 26 March 2020).

Borisova, V. 2018. 'Essential Characteristics And Market of the Creative Industries' Product' *Economic Alternatives* 1: 113–132, https://www.unwe.bg/uploads/Alternatives/8_Borisova_Alternativi_english_broi_1_2018.pdf) (accessed on 20 March 2020).

Burt, Roger and Colin Davies. 2018. 'Software: intellectual property and artificial intelligence' in Brown, Abbe E.L. and Charlotte Waelde: *Research Handbook on Intellectual Property and Creative Industries*. Edward Elgar Publishing.

Cauffman, Caroline. 2018. 'Robo-liability: The European Union in search of the best way to deal with liability for damage caused by artificial intelligence'. *Maastricht Journal of European and Comparative Law* 25 (5): 527-532, (https://doi.org/10.1177/1023263X18812333) (accessed on 5 February 2020).

Chopra, Samir. 2018. 'End intellectual property'. *Aeon*, (https://aeon.co/essays/the-idea-of-intellectual-property-is-nonsensical-and-pernicious) (accessed 18 March 2020).

Christie, Andrew F. 2011. 'Creativity and innovation: A legal perspective', in Creativity and Innovation in Business and Beyond: Social Science Perspectives and Policy Implications. Routledge.

Compendium of US Copyright Office Practices.

Dawkins. Marian S. 2012. *Through Our Eyes Only? The Search For Animal Consciousness*. Oxford University Press.

Dutfield, Graham. 2013. 'Collective Invention and Patent Law Individualism: Origins and Functions of the Inventor's Rights of Attribution'. *The WIPO Journal: Analysis of Intellectual Property Issues* 5 (1): 25-34, (https://www.wipo.int/edocs/pubdocs/en/intproperty/wipo_journal/wipo_journal_5_1.pdf) (accessed on 8 February 2020).

European Council Directive 85/374/EEC, 25 July 1985.

European Patent Office (EPO). 2020. EPO publishes grounds for its decision to refuse two patent applications naming a machine as inventor", (https://www.epo.org/news-issues/news/2020/20200128.html) (accessed 10 February 2020).

Felix Oberholzer–Gee and Koleman Strumpf. 2010.,'File Sharing and Copyright'. *Innovation Policy and the Economy* 10 (1): 19–55, (http://dx.doi.org/10.1086/605852) (accessed on 18 March 2020).

Firth-Butterfield, Kay and Kay Chae. 2018. 'Artificial Intelligence Collides With Patent Law'. Center for the Fourth Industrial Revolution, World Economic Forum (http://www3.weforum.org/docs/WEF_48540_WP_End_of_Innovation_Protecting_Patent_Law.pdf) (accessed on 29 March 2020).

Fraser, Erica. 2016. 'Computers as Inventors ---- Legal and Policy Implications of Artificial Intelligence on Patent Law'. *SCRIPTed* 13 (3): 305, (https://script-ed.org/article/computers-as-inventors-legal-and-policy-implications-of-artificial-intelligence-on-patent-law/) (accessed on 20 February 2020).

Gaut, Berys. 2010. 'The Philosophy of Creativity'. *Philosophy Compass* 5 (12): 1034-1046, (https://www.sfu.ca/~kathleea/docs/The%20Philosophy%20of%20Creativity%20-%20Gaut.pdf) (accessed on 5 February 2020).

Gneezy, U., Stephan Meier and Pedro Rey-Biel. 2011. 'When and Why Incentives (Don't) Work to Modify Behaviour'. *Journal of Economic Perspectives* 25(4) 191–210, (https://rady.ucsd.edu/faculty/directory/gneezy/pub/docs/jep_published.pdf) (accessed on 10 April 2020).

Government of United Kingdom. 2006. *Gowers Review of Intellectual Property*, (https://www.gov.uk/government/publications/gowers-review-of-intellectual-property) (accessed on 8 February 2020).

Guadamuz, A. 2017a. 'AI and copyright'. *WIPO Magazine*, (https://www.wipo.int/wipo_magazine/en/2017/05/article_0003.html.) (accessed on 22 February 2020).

Guadamuz, A. 2017b. 'Do androids dream of electric copyright? Comparative analysis of originality in artificial intelligence generated works'. *Intellectual Property Quarterly* 2, (https://ssrn.com/abstract=2981304) (accessed on 21 February 2020).

Hu, Cherie. 2019. 'Is AI-generated music worth anything' *The Industry Observer – The Brag* (https://theindustryobserver.thebrag.com/is-ai-generated-music-worth-anything/) (accessed on 2 April 2020).

Hughes, Justin. 1988. 'The Philosophy of Intellectual Property'. *Georgetown Law Journal*, (http://www.justinhughes.net/docs/a-ip01.pdf) (accessed on 20 February 2020).

Iglesias, M., Sharon Shamuilia and Amanda Anderberg, A. 2019. *Intellectual Property and Artificial Intelligence: A literature review*, Publications Office of the European Union, Luxembourg, (https://ec.europa.eu/jrc/en/publication/intellectual-property-and-artificial-intelligence-literature-review) (accessed on 1 March 2020).

Imagination Engines Inc. 'IEI's Patented Creativity Machine Paradigm'. Home of the Creativity Machine, (http://imagination-engines.com/iei_cm.php) (accessed on 8 February 2020).

Imagination Engines Inc. 'What is DABUS?'. *Home of the Creativity Machine*, (http://imagination-engines.com/iei_dabus.php) (accessed on 8 February 2020).

Indian Copyright Act, 1957.

Indian Patents Act, 1970.

Jehan, Robert. 2019. 'Should an AI system be credited as an inventor'. *The Artificial Inventor Project*, (http://artificialinventor.com/should-an-ai-system-be-credited-as-an-inventor-robert-jehan/) (accessed on 15 February 2020).

Kaplan, Jerry. 2016. *Artificial Intelligence: What Everyone Need To Know*. Oxford University Press.

Karpathy, Andrej. 2015. 'The Unreasonable Effectiveness of Recurrent Neural Networks' *Andrej Karpathy blog* (http://karpathy.github.io/2015/05/21/rnn-effectiveness/) (accessed 10 March 2020).

Keats, Jonathan. 2006. 'John Koza Has Built an Invention Machine'. *Popular Science*, (https://www.popsci.com/scitech/article/2006-04/john-koza-has-built-invention-machine/) (accessed on 12 February 2020).

Knorr, Eric. 2001. 'Origin of the Patents'. *MIT Technology Review*, (https://www.technologyreview.com/2001/08/03/102048/origin-of-the-patents/) (accessed on 10 March 2020).

Landes, William M. and Richard A. Posner. 2003. *The Economic Structure of Intellectual Property Law*. The Belknap Press of Harvard University.

Lauber-Rönsberg, A., Sven Hetmank. 2019. 'The concept of authorship and inventorship under pressure: Does artificial intelligence shift paradigms?'. *Journal of Intellectual Property Law & Practice* 14 (7), (https://doi.org/10.1093/jiplp/jpz061) (accessed on 5 April 2020).

Lester, S. and Huan Zhu. 2019. 'Rethinking the Length of Patent Terms'. *American University International Law Review* 34 (4): 787-886 (https://www.cato.org/sites/cato.org/files/2020-02/lester-zhu-auilr-v34n4.pdf) (accessed on 30 March 2020).

Lindblom, Charles E. 2016. 'The Science of Muddling Through', in Lodge Martin et al. (ed.) in *The Oxford Handbook of Classics in Public Policy and Administration*, (https://www.oxfordhandbooks.com/view/10.1093/oxfordhb/9780199646135.001.0001/oxfordhb-9780199646135-e-33) (accessed on 10 April 2020).

Lukose, Lisa P. 2013. 'Minors' Rights Under Intellectual Property Laws: A Myth or Reality?'. *Journal of Intellectual Property Rights* 18: 174-180, (http://nopr.niscair.res.in/bitstream/123456789/16398/1/JIPR%2018%282%29%20174-180.pdf) (accessed on 20 March 2020).

MacCauley, Thomas B. 1841. *Copyright – A Speech Delivered in the House of Commons*. House of Commons of the United Kingdom, (https://www.gutenberg.org/files/2170/2170-h/2170-h.htm#link2H_4_0018) (accessed on 6 February 2020).

Maher, Chauncey. 2017. *Plant Minds: A Philosophical Defense*. Routledge Focus on Philosophy

Marks, Paul. 2015. 'Eureka Machines'. *New Scientist* 227, (https://doi.org/10.1016/S0262-4079(15)31079-4) (accessed on 4 March 2020).

McLaughlin, M. 2018. 'Computer-Generated Inventions'. *Social Science Research Network*, (https://papers.ssrn.com/sol3/papers.cfm?abstract_id=3097822) (accessed on 28 February 2020).

Michaux, B. 2018. 'Singularité technologique, singularité humaine et droit d'auteur', in Laws, Norms and Freedoms in Cyberspace = Droit, normes et libertés dans le cybermonde: liber amicorum Yves Poullet: 401-416. (Collection du CRIDS; No. 43). Bruxelles: Larcier.

Nurton James. 'EPO and UKIPO Refuse AI-Invented Patent Applications'. *IPWatchdog*, (https://www.ipwatchdog.com/2020/01/07/epo-ukipo-refuse-ai-invented-patent-applications/id=117648/) (accessed 20 February 2020).

Okediji, Ruth L. 2018. 'Copyright Markets and Copyright in the Fourth Industrial Era: Reconfiguring the Public Benefit for a Digital Trade Economy'.

*International Centre for Trade and Sustainable Development* Issue Paper No. 43.

Olson, David S. 2009. 'Taking the Utilitarian Basis for Patent Law Seriously: The Case for Restricting Patentable Subject Matter'. *Temple Law Review* 82: 181–240, (https://lawdigitalcommons.bc.edu/lsfp/274/) (accessed on 6 February 2020).

Perry, Mark and Thomas Margoni. 2010. 'From music tracks to Google maps: Who owns computer-generated works'. *Computer Law & Security Review* 26 (6): 621-629, (https://core.ac.uk/download/pdf/46560343.pdf) (accessed on 1 March 2020).

Plotkin, Robert. 2009. *The Gene in the Machine: How Computer-Automated Inventing is Revolutionizing Law & Business*. Stanford University Press.

Pollock R. 2008. 'Innovation and Imitation with and without Intellectual Property Rights'. *Cambridge University*, (https://mpra.ub.uni-muenchen.de/5025/) (accessed on 12 March 2020).

Posner, Richard. 2012. 'Do patent and copyright law restrict competition and creativity excessively'. *The Becker-Posner Blog*, (https://www.becker-posner-blog.com/2012/09/do-patent-and-copyright-law-restrict-competition-and-creativity-excessively-posner.html) (accessed on 12 March 2020).

Ramalho, Ana. 2017. 'Will robots rule the artistic world? A proposed model for the legal status of creations by artificially intelligent systems. *Journal of Internet Law* 21 (1): 12–25.

Ravid, Shlomit Y. and Xiaoqiong (Jackie) Liu. 2018. 'When Artificial Intelligence Systems Produce Inventions: An Alternative Model for Patent Law at the 3A Era'. *Cardozo Law Review*, (http://cardozolawreview.com/wp-content/uploads/2018/08/RAVID.LIU_.39.6.5-1.pdf) (accessed on 18 February 2020).

Ross D. King et al. 2009. 'The Robot Scientist Adam'. *Computer* 42: 46-54, (https://doi.ieeecomputersociety.org/10.1109/MC.2009.270) (accessed on 2 April 2020).

Russell, Stuart and Peter Norvig. 2013. *Artificial Intelligence: A Modern Approach*. (3rd edition), (https://www.cin.ufpe.br/~tfl2/artificial-intelligence-modern-approach.9780131038059.25368.pdf) (accessed on 20 February 2020).

Schönberger, D. 2018. 'Deep Copyright: Up – And Downstream Questions Related to Artificial Intelligence (AI) and Machine Learning (ML)', in De Werra Jacques (ed.): *Droit d'auteur 4.0/Copyright 4.0*. Schulthess Editions Romandes, pp. 145-173, (https://ssrn.com/abstract=3098315) (accessed on 1 April 2020).

Seeber, Monica and Richard Balkwill. *Managing Intellectual Property in the Book Publishing Industry: A business-oriented information booklet*. Creative Industries – Booklet No. 1, World Intellectual Property Organization, (https://www.wipo.int/publications/en/details.jsp?id=255&plang=EN) (accessed on 10 February 2020).

Shemtov, 2019. 'A study on inventorship in inventions involving AI activity'. European Patent Office, (http://documents.epo.org/projects/babylon/

eponet.nsf/0/3918F57B010A3540C125841900280653/$File/Concept_of_Inventorship_in_Inventions_involving_AI_Activity_en.pdf) (accessed on 28 March 2020).

Sherer, F.M. 1972. 'Nordhaus' Theory of Optimal Patent Life: A Geometric Reinterpretation'. *The American Economic Review* 62 (3): 422-427, (https://www.jstor.org/stable/1803388) (accessed 10 February 2020).

Soni, Pankaj and Kartikey Vikrant Singh. 2019. 'Are we ready for AI disruption? An Indian patent law perspective'. *Remfry & Sagar*, (https://www.remfry.com/wp-content/uploads/2018/10/Print-PDF.pdf) (accessed on 10 March 2020).

Stilgoe, J., Richard Owen and Phil Macnaghten. 2013. 'Developing a framework for responsible innovation'. *Research Policy (Elsevier)* 42 (9): 1568-1580, (https://doi.org/10.1016/j.respol.2013.05.008) (accessed on 2 February 2020).

Stilgoe, Jack. 2018. 'Machine learning, social learning, and the governance of self-driving cars'. *Social Studies of Science* (https://doi.org/10.1177/0306312717741687) (accessed on 10 February 2020).

Tapscott, Rebecca. 2020. 'USPTO Shoots Down DABUS' Bid for Inventorship'. *IPWatchdog*, (https://www.ipwatchdog.com/2020/05/04/uspto-shoots-dabus-bid-inventorship/id=121284/) (accessed 6 May 2020).

Tata Consultancy Services and Confederation of Indian Industry. 2019.' Understanding the Dynamics of Artificial Intelligence in Intellectual Property', (https://www.tcs.com/content/dam/tcs/pdf/discover-tcs/about-us/press-release/understanding-dynamics-artificial-intelligence-intellectual-property.pdf) (accessed on 2 April 2020).

The Next Rembrandt. 2016. 'The Next Rembrandt: Data's new leading edge role in creativity' (https://thenextrembrandt.pr.co/130454-the-next-rembrandt) (accessed 10 March 2020).

Thurow, Lester C. 1997. 'Needed: A New System of Intellectual Property Rights'. *Harvard Business Review* (https://hbr.org/1997/09/needed-a-new-system-of-intellectual-property-rights) (accessed on 10 February 2020).

Torrance, Andrew W. and Bill Tomlinson, 2009. 'Patents and the Regress of Useful Arts'. *Columbia Science and Technology Law Review* 10, (https://ssrn.com/abstract=1411328) (accessed on 15 March 2020).

UK Copyright Designs and Patents Act, 1988.

UK Patents Act, 1977.

United Kingdom Intellectual Property Office (UKIPO). 2017. *Formalities Manual* ch 3 (updated on 1 February 2020), (https://www.gov.uk/guidance/formalities-manual-online-version/chapter-3-the-inventor) (accessed on 10 February 2020).

United States Patent and Trademark Office, (USPTO), Department of Commerce. 2019. *Request for Comments on Intellectual Property Protection for Artificial Intelligence Innovation*, (https://www.govinfo.gov/content/pkg/FR-2019-10-30/pdf/2019-23638.pdf) (accessed on 2 February 2020).

Varian, Hal R. 2010. 'Computer Mediated Transactions'. *American Economic Review: Papers and Proceedings* 100: 1–10 (http://www.aeaweb.org/articles.php?doi=10.1257/aer.100.2.1) (accessed on 10 February 2020).

Veale, Tony and F. Amilcar Cardoso. 2019. *Computational Creativity: The Philosophy and Engineering of Autonomously Creative Systems* Springer Nature Switzerland AG.

Vertinsky, Liza and Todd M. Rice. 2002. 'Thinking About Thinking Machines: Implications of Machine Inventors for Patent Law'. *Boston University Journal of Science and Technology Law* 8: 574–613, (https://www.bu.edu/law/journals-archive/scitech/volume82/vertinsky&rice.pdf) (accessed on 25 March 2020).

Viscusi, William K. and Michael J. 1993. 'Moore, Product Liability, Research and Development, and Innovation' *Journal of Political Economy* 101 (1): 161-184, (https://www.jstor.org/stable/2138678) (accessed on 15 March 2020).

Wachowicz, Marcos and Goncalves Lukas Reuthes. 2019. *Artificial Intelligence and Creativity: New Concepts in Intellectual Property*. Curitiba: GEDAI Publications.

Walmsley. C. 2011. 'Flashes of Genius, Toiled Experimentation, and Now Artificial Creation: A Case for Inventive Process Disclosures' *The George Washington University* LLM Dissertation, (http://etd.gelman.gwu.edu/etd_11165/11165.pdf) (accessed 12 February 2020).

World Intellectual Property Organization (WIPO). 'What is Intellectual Property?', (https://www.wipo.int/edocs/pubdocs/en/intproperty/450/wipo_pub_450.pdf) (accessed 20 February 2020).

World Intellectual Property Organization (WIPO). 2019(a). *Draft Issues Paper on Intellectual Property Policy and Artificial Intelligence*, (https://www.wipo.int/meetings/en/doc_details.jsp?doc_id=470053) (accessed on 2 February 2020).

World Intellectual Property Organization (WIPO). 2019(b). *Background Documents on Patents and Emerging Technologies*. Geneva: Standing Committee on the Law of Patents, Thirtieth Session, WIPO, (https://www.wipo.int/edocs/mdocs/scp/en/scp_30/scp_30_5.pdf) (accessed on 10 March 2020).

Youn et al. 2015. 'Invention as a combinatorial process: evidence from US patents'. *Journal of the Royal Society Interface* 12 (106), (https://doi.org/10.1098/rsif.2015.0272) (accessed on 10 February 2020).